\documentclass[12pt]{article}

\usepackage{amssymb}

\begin{document}

\begin{titlepage}

\begin{flushright}
%arXiv:XXXX.XXXX
\end{flushright}
\vskip 2.5cm

\begin{center}
{\Large \bf No Vacuum Cerenkov Radiation Losses in the Timelike
Lorentz-Violating Chern-Simons Theory}
\end{center}

\vspace{1ex}

\begin{center}
{\large Karl Schober, Brett Altschul\footnote{{\tt baltschu@physics.sc.edu}}}

\vspace{5mm}
{\sl Department of Physics and Astronomy} \\
{\sl University of South Carolina} \\
{\sl Columbia, SC 29208} \\
\end{center}

\vspace{2.5ex}

\medskip

\centerline {\bf Abstract}

\bigskip

In a Lorentz- and CPT-violating modification of electrodynamics that
includes a timelike Chern-Simons term, there are no energy losss through
vacuum Cerenkov radiation. A charge moving with a constant
velocity does not lose energy, because of an unusual cancellation. Higher
frequency modes of the electromagnetic field carry away positive
energy, but lower frequency modes carry away a compensating negative
amount of energy.
% This is possible because the energy density is not bounded from below.

\bigskip

\end{titlepage}

\newpage

Recent years have seen a growth in interest in the possibility that Lorentz
and CPT symmetries might be violated at the most fundamental levels of
physics. These symmetries are ordinarily considered basic building blocks
for the laws of nature, but they might not hold in theories of quantum
gravity. In fact, many of the theories that have been proposed as
candidate theories of quantum gravity may have regimes with Lorentz
symmetry violation. Conversely, if any violations of Lorentz or CPT
symmetries are ever uncovered experimentally, that will be a discovery
of paramount importance, opening up new windows on the structure of
fundamental physics.

There are also other motivations for studying exotic symmetry-breaking
theories. Even if Lorentz violation and CPT violation are not found to exist in
nature, by studying theories with these kinds of symmetry breakings, we may
gain new insights about the basic structure of quantum field theories.
Such studies may be conducted using effective quantum field theory. There is
an effective theory called the standard model extension (SME) that 
has been developed to deal with Lorentz and CPT violation in fairly general
fashion. The action for the SME contains all operators that can be
constructed using standard model fields~\cite{ref-kost1,ref-kost2}.
Much more general operators are possible in the SME than in the standard
model itself, because the operators are not constrained by Lorentz
invariance. While the full SME obviously contains an infinite number of
possible operators, the minimal SME, which includes only local, gauge
invariant, and superficially renormalizable operators, provides a tractable
test theory for exploring the experimental implications of Lorentz and CPT
violations.

This paper will explore a particularly interesting process that can occur
in many Lorentz-violating modifications of quantum electrodynamics---vacuum
Cerenkov radiation, in which a charged particle moving
with constant velocity through the vacuum emits radiation. Despite all
appearances to the contrary (and expectations 
based on the usual threshold condition for Cerenkov emission),
charges in uniform motion in the theory we shall
consider do not lose energy through Cerenkov radiation.

Although it is possible to have Lorentz violation in both the electromagnetic
and matter sectors, we shall concentrate on one particular form of Lorentz
violation that is purely electromagnetic. The term we shall consider---the
timelike electromagnetic Chern-Simons term---is one of the most interesting
terms that appear in the SME action. This Chern-Simons term introduces
a dimensional scale into pure electrodynamics. It leads to a screening of
magnetostatic fields and differing dispersion relations for right- and
left-handed electromagnetic waves. This kind of birefringence would have a
distinctive experimental signature, which has been searched for and not seen,
even for photons that have traversed cosmological
distances~\cite{ref-carroll1,ref-mewes5}.
This leads to exceedingly strong experimental constraints on the coefficient
of the Chern-Simons operator.

However, in spite of these tight empirical bounds,
the Chern-Simons term remains extremely interesting; understanding the behavior of
quantum electrodynamics with an added Chern-Simons term may reveal new insights
about how quantum field theories may behave. For example,
the Chern-Simons term in the SME Lagrange density is not gauge invariant;
however, since the term changes by a total derivative under a gauge transformation,
the integrated action associated with the Chern-Simons term is gauge invariant.
This subtlety makes determining the radiative corrections to the Chern-Simons term
an extremely tricky problem, which led to some significant controversy~\cite{ref-coleman,
ref-jackiw1,ref-victoria1,ref-chung1,ref-andrianov,ref-altschul1}.

Naively, vacuum Cerenkov radiation looks like it should be allowed in the
Chern-Simons theory. Ordinary Cerenkov
radiation only occurs in a medium, because in a Lorentz-invariant vacuum, a process
like $e^{-}\rightarrow e^{-}+\gamma$ is disallowed by energy-momentum
conservation. However, it is common in Lorentz-violating theories for processes
that are normally kinematically forbidden to become allowed. The naive
condition for Cerenkov emission---that the speed of a moving charge exceeds the
phase speed of light propagating in the same direction---can be readily
satisfied in the Chern-Simons theory. However, we shall see that whether
Cerenkov losses actually occur in this theory is a much more subtle question.

The question of vacuum Cerenkov radiation in the SME has been studied before,
in several different SME parameter regimes. The previous analyses
have used a variety of techniques. Some studies have applied Green's
functions~\cite{ref-lehnert2} and the related Feynman diagram
techniques directly~\cite{ref-kaufhold}. Other analyses used 
coordinate transformations to move the Lorentz violation
into the charged matter sector; this can be combined with the observation that when
the electromagnetic sector is conventional, Cerenkov radiation
occurs exactly if a charge's speed $v$ exceeds 1~\cite{ref-altschul9,ref-altschul12}.

A fully comprehensive picture of Cerenkov radiation in the minimal SME
has still not yet been developed, however. This work provides an exact
(and highly unexpected) solution to the problem in the timelike
Chern-Simons theory---which was among the first renormalizable Lorentz-violating
theories to be studied. Rather than working with the Green's functions and
dispersion relations for the Lorentz-violating theory, this analysis will work
directly with the modified Maxwell's equations. We shall develop a systematic
way to find electric and magnetic fields of a uniformly moving charge, perturbatively as
a function of the Lorentz violation coefficient and the particle speed. From these
fields, the radiation losses may be calculated directly.
This technique was first applied in~\cite{ref-altschul36} up to ${\cal O}(v^{2})$,
for which only a determination of a single modified field at first order in the
Lorentz violation was needed. Here we shall demonstrate how the same
kind of calculations may be performed to all orders.

The Lagrange density for the electromagnetic sector of the minimal SME is
\begin{equation}
{\cal L}=-\frac{1}{4}F^{\mu\nu}F_{\mu\nu}
-\frac{1}{4}k_{F}^{\mu\nu\rho\sigma}F_{\mu\nu}F_{\rho\sigma}
+\frac{1}{2}k_{AF}^{\mu}\epsilon_{\mu\nu\rho\sigma}F^{\nu\rho}A^{\sigma}
-j^{\mu}A_{\mu}.
\end{equation}
The $k_{F}$ term, which is even under CPT, has been extensively studied. The
focus here will be on the $k_{AF}$ term; this is the Chern-Simons term, which is
odd under CPT. We shall specifically be interested in a timelike $k_{AF}$ term,
which, in an appropriate frame, takes the purely timelike form
$k_{AF}^{\mu}=(k,\vec{0}\,)$.

There is an obvious difficultly with this theory
when we look at the dispersion relations for circularly polarized plane waves. The
dispersion relation is $\omega_{\pm}^{2}=p(p\mp 2k)$, for modes of helicity $\pm 1$.
At the longest wavelengths, with $p<|2k|$, one of the two modes apparently has an
imaginary frequency; as we will see, this is related to the fact that the energy
is not bounded below. The existences of imaginary frequencies and unboundedly negative
energies are, in turn, tied to the existence of runaway solutions, in which the field grows
without bound. It is possible to avoid the runaway solutions by using an acausal
Green's function~\cite{ref-carroll1}. However, there is a cost, lying
precisely in the acausality;
charged particles may begin to radiate before they actually start to move. Identifying
the correct behavior of a theory with a timelike Chern-Simons term thus becomes a
tricky problem. Fortunately, however, this problem does not exist when we study the
Cerenkov radiation from a charge that moves with a perfectly constant velocity.

To understand the radiation of energy in a modified electromagnetic theory, we must
look at the energy-momentum tensor. For an arbitrary $k_{AF}$, this tensor
takes the form~\cite{ref-carroll1}
\begin{equation}
\Theta^{\mu\nu}=-F^{\mu\alpha}F^{\nu}\,_{\alpha}+\frac{1}{4}g^{\mu\nu}
F^{\alpha\beta}F_{\alpha\beta}-\frac{1}{2}k_{AF}^{\nu}\epsilon^{\mu\alpha\beta\gamma}
F_{\beta\gamma}A_{\alpha}.
\end{equation}
For the purely timely $k_{AF}$, the energy density
($\Theta^{00}$), momentum density ($\Theta^{0j}$), and energy flux ($\Theta^{j0}$) are
\begin{eqnarray}
{\cal E} & = & \frac{1}{2}\vec{E}^{2}+\frac{1}{2}\vec{B}^{2}-
k\vec{B}\cdot\vec{A} \\
\vec{{\cal P}} & = & \vec{E}\times\vec{B} \\
\vec{S} & = & \vec{E}\times\vec{B}-kA_{0}\vec{B}+k\vec{A}\times\vec{E}.
\end{eqnarray}
The asymmetry of $\Theta^{\mu\nu}$ is a general feature of Lorentz-violating
theories. However, other unusual features are specific to the theory with
$k_{AF}$. The energy density is not gauge invariant, although the total
energy, integrated over all space, is a gauge-invariant quantity. Moreover, as
previously noted, ${\cal E}$ is not bounded below, and this feature will turn out
to plan a crucial role in understanding the absence of Cerenkov radiation.

To study Cerenkov radiation, we will look at the fields produced by a charged
particle $q$ moving along the trajectory $\vec{r}(t)=vt\hat{z}$, taken at
time $t=0$. A particle
emitting Cerenkov radiation would of be subject to recoil; however, we shall
neglect any effects of recoil here. (Any radiation that is directly a result
of recoil effects is not really Cerenkov radiation, since it is caused by
a secondary acceleration.) For a particle in completely steady motion, all the
fields it produces must similarly be moving along in the $z$-direction with
speed $v$, and this will provide a tremendous simplification of the time
derivatives that appear in Maxwell's equations.

We write the electric and magnetic fields in this scenario as series
\begin{eqnarray}
\vec{E} & = & \sum_{m=0}^{\infty}\sum_{n=0}^{\infty}\vec{E}^{(m,n)} \\
\vec{B} & = & \sum_{m=0}^{\infty}\sum_{n=1}^{\infty}\vec{B}^{(m,n)}.
\end{eqnarray}
In each sum, a term with superscripts $(m,n)$ is proportional to
$k^{m}v^{n}$.
These power seies forms will allow us to pick out the terms that can contribute
to our expression for the radiated energy. The usual terms are the
$\vec{E}^{(0,n)}$ with $n$ even and $\vec{B}^{(0,n)}$ with $n$ odd.

The only modification of Maxwell's equations is to the Ampere-Maxwell law;
the modified version is
\begin{equation}
\vec{\nabla}\times\vec{B}=\frac{\partial\vec{E}}{\partial t}+2k\vec{B}+\vec{J}.
\end{equation}
The other three equations are just as usual. The time
derivatives of the fields simplify considerably, since we know that the
steady-state field
profiles must be moving in the $z$-direction uniformly in time. By considering
the steady-state scenario, we avoid having to deal with the question of whether
an acausal Green's function in required. A field (either $\vec{E}$ or $\vec{B}$)
must have the form $\vec{W}(\vec{r},t)=\vec{W}(\vec{r}-v\hat{z}t,0)$,
and this means that
any time derivative $\frac{\partial}{\partial t}$ acting on $\vec{E}$ or $\vec{B}$
may be replaced with $-v\frac{\partial}{\partial z}$.

The modified fields may thus be generated iteratively, according to
\begin{eqnarray}
\label{eq-Ecurl}
\vec{\nabla}\times\vec{E}^{(m,n)} & = & v\frac{\partial\vec{B}^{(m,n-1)}}{\partial z} \\
\vec{\nabla}\cdot\vec{E}^{(m,n)} & = & 0 \\
\label{eq-Bcurl}
\vec{\nabla}\times\vec{B}^{(m,n)} & = & -v\frac{\partial\vec{E}^{(m,n-1)}}
{\partial z}+2k\vec{B}^{(m-1,n)} \\
\label{eq-Bdiv}
\vec{\nabla}\cdot\vec{E}^{(m,n)} & = & 0,
\end{eqnarray}
starting with the Lorentz-invariant fields $\vec{E}^{(0,n)}$  and $\vec{B}^{(0,n)}$.
Successively solving these equations provides terms with increasing $m+n$. Note that
no terms beyond $m=2$ can contribute to the energy loss; for dimensional reasons,
the total power radiated must have mass dimension 2, and there is no other
dimensional scale apart from $k$ in the problem.

Understanding the general properties of the $\vec{E}^{(m,n)}$ and
$\vec{B}^{(m,n)}$ is crucially important. The standard fields $\vec{E}^{(0,n)}$
and $\vec{B}^{(0,n)}$ involve only even powers of $v$ for the electric terms and
odd powers of $v$ for the magnetic terms. Looking at the equations
(\ref{eq-Ecurl}--\ref{eq-Bdiv}), it is evident that $\vec{E}$ remains
an even function of $v$ and $\vec{B}$ an odd function to all orders. However,
there are no similar conditions on the parity of the solutions with respect to $k$.

Geometrically, the $k$-dependent $\vec{E}^{(m,n)}$ and $\vec{B}^{(m,n)}$
terms may take two forms; they may be either azimuthal or toroidal. An azimuthal
field is divergenceless and points in the $\hat{\phi}$-direction, and its magnitude
is independent of $\phi$; the $\vec{B}^{(0,n)}$ all have this form. A toroidal field
is likewise divergenceless, and it points entirely in the $\hat{r}$- and
$\hat{\theta}$-directions; and again, these components are independent of $\phi$.
Whichever geometric structure a
given term $\vec{W}$ has, its derivative $\partial{\vec{W}}/\partial z$
will have the same structure. (The $z$-derivative of a divergenceless field remains
divergenceless; nor will the derivative
introduce any explicit dependence on $\phi$; nor does taking the derivative
mix the $\{\hat{r},\hat{\theta}\}$ and $\{\hat{\phi}\}$ spaces of unit vectors.)

A key point is that if the sources for a given term [that is, the fields appearing on
the right-hand sides of (\ref{eq-Ecurl}) and (\ref{eq-Bcurl})] are toroidal, then the
terms they generate (on the left) are azimuthal, and vice versa. These are both standard
facts. The most elementary examples of these statements are that the magnetic field
of a circular loop of current is toroidal, and the magnetic field of a toroid of uniform
cross section wound with wire is azimuthal. The general statements can, in fact, be derived
simply by taking superpositions of sources with the geometries appearing those two examples.
Since the $\vec{B}^{(0,n)}$ (from which all the $k$-dependent terms are ultimately
derived) are azimuthal, it follows (by trivial induction) that a field has toroidal
geometry whenever $m+n$ is even and azimuthal geometry when $m+n$ is odd. No term has a
mixture of the two geometry types.

The solutions of the equations (\ref{eq-Ecurl}--\ref{eq-Bdiv}) may be found by elementary
means. The problem is easier when the source terms are toroidal, so that the term they
generate is azimuthal. The magnitude of the azimuthal component may be found using
a pseudo-Amperean technique. Consider a circle $C$ with radius $\rho$, lying parallel to the
$xy$-plane and with its center at $(0,0,z)$. For concreteness, we will look at how
$\vec{B}^{(2,1)}$ is generated from the
$\vec{B}^{(1,1)}=\frac{kqv}{4\pi r}(2\cos\theta\,\hat{r}-\sin\theta\,\hat{\theta})$
previously calculated in~\cite{ref-altschul36}.
We have
\begin{eqnarray}
\int_{C}d\vec{l}\cdot\vec{B}^{(2,1)}=2\pi\rho B^{(2,1)}_{\phi} & = & \int_{0}^{\rho}
(2\pi \rho'\,d\rho')\vec{B}^{(1,1)}\cdot\hat{z} \\
& = & \frac{kqv}{2}\int_{0}^{\rho}d\rho'\frac{\rho'}{\sqrt{\rho'^{2}+z^{2}}}
\left(1+\frac{z^{2}}{\rho'^{2}+z^{2}}\right) \\
& = & \frac{kqv}{2}\frac{\rho^{2}}{\sqrt{\rho^{2}+z^{2}}} \\
\vec{B}^{(2,1)} & = & \frac{kqv}{4\pi}\sin\theta\,\hat{\phi}
\end{eqnarray}
This has a singularity at the origin, but so does its source, and so indeed do
all the fields.

It will also be critical to understand whether the components of the
$\vec{E}^{(m,n)}$ and $\vec{B}^{(m,n)}$ are odd or even functions of $\cos\theta$ (or
equivalently of $z$). This will be referred to as the $z$-parity. Ultimately,
the $z$-parity will be important because if $\vec{S}\cdot\vec{\hat{r}}$ is an
odd function of $\cos\theta$, there will be no net outflow of energy.

In the example calculation just considered, with $\vec{B}^{(1,1)}$ generating
$\vec{B}^{(2,1)}$, the source could be written as
$\vec{B}^{(1,1)}=f(r)[O(\theta)\,\hat{r}+E(\theta)\,\hat{\theta}]$, where
$f(r)$ contains the radial dependence, $O(\theta)$ is an odd function of $\cos\theta$
(although, obviously, not an odd function of $\theta$ itself), and $E(\theta)$ is
an even function of $\cos\theta$. Then
$\vec{B}^{(1,1)}\cdot\hat{z}=f(r)[O(\theta)\cos\theta-E(\theta)\sin\theta]$ is
strictly an even function of $\cos\theta$. The flux of this field through a circular
loop centered at $(0,0,z)$ is the same as it would be if the loop were centered at
$(0,0,-z)$; this makes the azimuthal $\vec{B}^{(2,1)}$ an even function of $\cos\theta$.
Obviously, this generalizes to other sources with the same $z$-parity structure; and
a source term with the opposite $z$-parity structure [proportional to $E(\theta)\,\hat{r}
+O(\theta)\,\hat{\theta}$] will generate an new azimuthal field term
with odd $z$-parity.

In addition to $\vec{B}^{(m,n)}$ serving directly as a source (via the novel term in 
Maxwell's equations), there are also the source terms in the usual electromagnetic
induction expressions, which involve derivatives. The $z$-derivative operator
\begin{equation}
\frac{\partial}{\partial z}=\cos\theta\frac{\partial}{\partial r}+
\frac{\sin^{2}\theta}{r}\frac{\partial}{\partial\cos\theta}
\end{equation}
reverses the $z$-parity of any field it acts on. Continuing with the example of a
$\vec{B}^{(1,1)}$ source, we have
\begin{equation}
\frac{\partial\vec{B}^{(1,1)}}{\partial z}=-\frac{kqv}{4\pi r^{2}}(3\cos^{2}\theta-1)\hat{r},
\end{equation}
which does indeed have the $E(\theta)\,\hat{r}+O(\theta)\,\hat{\theta}$ form
[albeit with $O(\theta)=0$].

The result of the preceding paragraphs' analysis is that
with a toroidal $\vec{B}^{(m,n)}$---having
odd $z$-parity in the $r$-component and even $z$-parity in the $\theta$-component---the
new field terms generated through (\ref{eq-Ecurl}) and (\ref{eq-Bcurl}) will be an
azimuthal $\vec{B}^{(m+1,n)}$ with even $z$-parity and an 
azimuthal $\vec{E}^{(m,n+1)}$ with odd $z$-parity. For
a toroidal $\vec{E}^{(m,n)}$ with the opposite $z$-parities (odd for the $r$-component
and even for the $\theta$-component) the sole
new term generated will be an
azimuthal $\vec{B}^{(m,n+1)}$, again with even $z$-parity.

The determinations of the new toroidal field terms generated by an azimuthal source
term is slightly tricky. There is no pseudo-Amperean technique for simply
reducing the solution of the problem to a definite integral. However, the problem can be
reduced to that of solving a single ordinary differential equation. The key to this
reduction is noticing that the dependence of any field on $r$ is completely determined
by its dependence on $k$. For dimensional reasons alone, $k$ and $r$ will always appear
in a single combined factor of $\frac{k^{m}}{r^{2-m}}$. Then we may take, for example,
$\vec{E}^{(m,n)}$ to be $\vec{E}^{(m,n)}=\frac{1}{r^{2-m}}[X(\theta)\hat{r}+
Y(\theta)\hat\theta]$, so that the curl and divergence
equations reduce to
\begin{eqnarray}
\label{eq-curlreduced}
\vec{\nabla}\times\vec{E}^{(m,n)} & = & \frac{1}{r^{3-m}}[-X'(\theta)+(m-1)Y(\theta)]
\hat{\phi}\,=\,v\frac{\partial B_{\phi}^{(m,n-1)}}{\partial z} \\
\label{eq-divreduced}
\vec{\nabla}\cdot\vec{E}^{(m,n)} & = & \frac{1}{r^{3-m}}[mX(\theta)+
\cot\theta\,Y(\theta)+Y'(\theta)]\,=\,0.
\end{eqnarray}
The prime in $X'$ denotes the derivative with respect to the argument
$\theta$. Note that if $X$ and $Y$ are functions of well-defined
$z$-parity, (\ref{eq-divreduced}) implies that the $z$-parities for the
two component functions $X$ and $Y$ are opposite.

For $m\neq 1$, (\ref{eq-curlreduced}) may be used to eliminate $Y$. The
resulting second-order differential equation is
\begin{equation}
\label{eq-odeX}
X''+\cot\theta X'+m(m-1)X=vr^{3-m}\left[\cot\theta
\frac{\partial B_{\phi}^{(m,n-1)}}{\partial z}-
\sin\theta\frac{\partial}{\partial\cos\theta}
\frac{\partial B_{\phi}^{(m,n-1)}}{\partial z}\right],
\end{equation}
using $\frac{\partial}{\partial\theta}=-\sin\theta\frac{\partial}
{\partial\cos\theta}$. There is a particular solution of this
equation that is proportional to $k^{m}v^{n}$ and has no
singularities in $\theta$. In all cases that we have specifically
investigated, this solution could be found using the Ansatz that
$X(\theta)$ be a polynomial in $\cos\theta$. Then $Y$ follows from
(\ref{eq-curlreduced}). For example, the solution for
$\vec{E}^{(2,2)}$ [sourced by a $\vec{B}^{(2,1)}$] is
\begin{equation}
\vec{E}^{(2,2)}=-\frac{k^{2}qv^{2}}{4\pi}\left[\left(\frac{3}{2}
\cos^{2}\theta-\frac{1}{2}\right)\hat{r}-\sin\theta\cos\theta\,\hat{\theta}\right].
\end{equation}

However, the key general point [evident from inspection of~(\ref{eq-odeX})] is that
the nonsingular particular solution $X$ has the same
$z$-parity as $B_{\phi}^{(m,n-1)}$. As noted, $Y$ then has the opposite $z$-parity.
(For the previously excluded $m=1$
case, the results concerning the $z$-parity of $X$ and $Y$ are the
same. However, the calculations at that order are simpler; only solutions of
first order linear differential equations are required.)

Collecting these results, we have that an azimuthal $\vec{B}^{(m,n)}$ with
even $z$-parity produces a toroidal $\vec{B}^{(m+1,n)}$ with odd $z$-parity in
the $\hat{r}$ component and even $z$-parity in the $\hat{\theta}$ component,
as well as a toroidal $\vec{E}^{(m,n+1)}$ with even $z$-parity along
$\hat{r}$ and odd for $\hat{\theta}$. An azimuthal $\vec{E}^{(m,n)}$ with
odd $z$-parity produces a toroidal $\vec{B}^{(m,n+1)}$ with odd $z$-parity in
$\hat{r}$ and even $z$-parity in $\hat{\theta}$.

\begin{table}
\begin{center}
\begin{tabular}{|c|c|c|c|}
\hline
& & $z$-Parity & \\
Field Term & $\hat{r}$ & $\hat{\theta}$ & $\hat{\phi}$ \\
\hline
Toroidal $\vec{E}$ & $+$ & $-$ & $\emptyset$ \\
Azimuthal $\vec{E}$ & $\emptyset$ & $\emptyset$ & $-$ \\
Toroidal $\vec{B}$ & $-$ & $+$ & $\emptyset$ \\
Azimuthal $\vec{B}$ & $\emptyset$ & $\emptyset$ & $+$ \\
\hline
\end{tabular}
\caption{
\label{table-zparity}
$z$-parity values for the four different types of field terms:
$+$ and $-$ denote even and odd parity, respectively;
$\emptyset$ denotes that
the corresponding term is zero for fields of the indicated type.}
\end{center}
\end{table}

Since the $k$-independent fields $\vec{E}^{(0,n)}$ and
$\vec{B}^{(0,n)}$ have well-defined $z$-parities, the higher order
field terms must as well. The $z$-parity values for the possible
fields, as determined from the rules we have derived, are listed in
table~\ref{table-zparity}. Knowing these symmetry properties, it is
possible to demonstrate that there is no outgoing energy flux at spatial
infinity, up to any order in $v$.

The energy flux $\vec{S}$ is composed of the various fields and
potentials. A net outward power loss is $P=R^{2}\int d\Omega\,
\vec{S}\cdot\hat{r}$, with the integral taken as $R\rightarrow\infty$.
Note that while ${\cal E}$ and $\vec{S}$ are not gauge invariant,
the integral of $\vec{S}\cdot\hat{r}$ over the sphere at infinity is a
gauge invariant quantity. If the integrand $\vec{S}\cdot\hat{r}$ itself has
odd $z$-parity, then integral automatically vanishes.

A number of different components of $\vec{E}$ and $\vec{B}$ may
contribute to $\vec{S}\cdot\hat{r}$. The conventional term
$(\vec{E}\times\vec{B})\cdot\hat{r}$---also the
only term appearing in the momentum outflow---receives
contributions proportional to $E_{\theta}B_{\phi}$ and
$E_{\phi}B_{\theta}$; both of these combinations have odd $z$-parity.
The remaining two terms involve the potentials $A_{0}$ and $\vec{A}$. Since
the corresponding terms in $\vec{S}$ depend on $k$ explicitly, and the
final power $P$ must be proportional to $k^{2}$, any contributing terms
must involve either $\vec{E}^{(1,n)}$ or $\vec{B}^{(1,n)}$, as the
$\vec{E}^{(0,n)}\propto\hat{r}$ and $\vec{B}^{(0,n)}\propto\hat{\phi}$
conventional terms obviously point in the wrong directions to contribute.
The potential terms needed are thus only the conventional
\begin{eqnarray}
A_{0} & = & \frac{q}{4\pi r\sqrt{1-v^{2}\sin^{2}\theta}} \\
\vec{A} & = & \frac{qv\hat{z}}{4\pi r\sqrt{1-v^{2}\sin^{2}\theta}}.
\end{eqnarray}
This $A_{0}$ has even $z$-parity, and the components of $\vec{A}$ have
the same $z$-parities as a the components of a toroidal $\vec{B}$. The
combinations involving these potentials that can contribute to $\vec{S}\cdot\hat{r}$
are thus proportional to $A_{0}B_{r}$, $A_{\theta}E_{\phi}$, and $A_{\phi}E_{\theta}$.
Again, all of these bilinears have odd $z$-parity.

Strictly speaking, the condition that $P$ be proportional to $k^{2}$ only
applies if $P$ is finite. This is not an idle concern; in the CPT-even theory
with $k_{F}$, the rate of Cerenkov losses for a charge moving
with constant speed is in fact infinite, and only recoil corrections
or additional higher-dimensional operators make the actual power emitted
finite. However, unlike in the $k_{F}$ theory, in the Chern-Simons theory
there is a hard upper limit on the energies of the emitted photons, above
which radiation is kinematically disallowed. Moreover,
it is actually a straightforward matter to generalize our analysis of
the $z$-parity to the $k$-dependent potential terms. This is particularly
transparent for the vector potential, for which the source relations in
the Coulomb gauge are
\begin{eqnarray}
\vec{\nabla}\times\vec{A}^{(m,n)} & = & \vec{B}^{(m,n)} \\
\vec{\nabla}\cdot\vec{A}^{(m,n)} & = & 0,
\end{eqnarray}
very similar to (\ref{eq-Bcurl}--\ref{eq-Bdiv}). So the $\vec{A}$ terms
have the same $z$-parities as $\vec{B}$ terms of the same geometric types.
The structure of $A_{0}$ is slightly trickier; the lone source equation for
the scalar potential is
\begin{equation}
\vec{\nabla}A_{0}^{(m,n)}=-\vec{E}^{(m,n)}+v\frac{\partial\vec{A}^{(m,n-1)}}
{\partial z}.
\end{equation}
The solutions $A_{0}^{(m,n)}$ (which only exist for $m+n$ even) have even
$z$-parity---leading to $\vec{\nabla}A_{0}^{(m,n)}$ with even $z$-parity
in the $\hat{r}$-component and odd $z$-parity in $\hat{\theta}$.

Since all possible contributions to $\vec{S}\cdot\hat{r}$ have odd $z$-parity,
the power loss from the moving charge is identically $P=0$. While $\vec{S}$
itself does not vanish at spatial infinity, $\vec{S}$ on its own is not a
gauge invariant quantity and has no physical meaning. Only the integral
can be measured. However, even taking this into account, $P=0$ is a puzzling
result. The previous analysis~\cite{ref-altschul36} found that $P=0$ up to
${\cal O}(v^{2})$; however, a phase space estimate of the power emitted at
${\cal O}(v^{3})$ was nonzero. Yet that estimate was based on an analysis only of the
modes with $p>|2k|$. For any $v>0$, there are $p>|2k|$ modes of the electromagnetic
field with phase speeds slower than $v$, and on general principles,
it appears there should
be energy radiated into these modes. However, the phase space analysis is
inapplicable for the $p<|2k|$ modes, which do not have well-defined real
frequencies. The behavior of these long-wavelength modes is actually key
to understanding why the total power emitted vanishes. Since the energy
density in the Chern-Simons theory is unbounded below, these modes of the field
actually carry away negative energy that precisely cancels the positive energies
borne away from the moving charge by the shorter-wavelength modes. Remarkably,
in this case, the lack of a lower bound on the energy actually makes the theory
better behaved than naive expectations would indicate; for a charged particle
moving with an arbitrary velocity $v$ in the timelike Chern-Simons theory, there
is no energy loss to vacuum Cerenkov radiation---just like in conventional
electromagnetism.

We have established the primary result, that there are no Cerenkov energy losses
in this model. However, this is still considerably more to delve into. While
no energy is lost by a charge in uniform motion, propagating electromagnetic
modes are excited. The phase space estimates of emission spectrum outlined
in~\cite{ref-altschul12,ref-altschul36} ought to tell us something about the
spectrum, even if they certainly
do not tell the whole story. A detailed, mode-by-mode understanding of the
radiation modes would be interesting, although it is clearly a complicated
problem---not least because the gauge dependence of ${\cal E}$ may make it
impossible to uniquely identify the energy carried by a particular mode.
Alternatively, it might be possible to obtain closed form expressions for
$\vec{E}$ and $\vec{B}$ fields. Many of the terms we have evaluated
explicitly have components proportional either to $P_{\ell}(\cos\theta)$ or
$P_{\ell}(\cos\theta)\sin\theta$, in terms of Legendre polynomials. It
might be possible, using functional identities, to resum the power
series to get a complete solution for the fields. Either one of these avenues
would be quite interesting, but the present work has demonstrated
the most important result, that the Cerenkov power loss rate in
the timelike Chern-Simons theory is $P=0$.

\end{document}